\documentclass[prl,twocolumn,amssymb]{revtex4}
\usepackage{graphicx}% Include figure files
\usepackage{epstopdf}
\usepackage{dcolumn}% Align table columns on decimal point
\usepackage{bm}% bold math
\usepackage{wasysym}
\newcommand{\beq}{\begin{equation}}
\newcommand{\eeq}{\end{equation}}
\newcommand{\beqn}{\begin{eqnarray}}
\newcommand{\eeqn}{\end{eqnarray}}
\newcommand{\nn}{\nonumber\\}
\newcommand{\Eq}[1]{Eq.~(\ref{#1})}

\begin{document}
\preprint{arXiv:1006.3794}
\title{Holographic metals and the fractionalized Fermi liquid}

\author{Subir Sachdev}
\affiliation{Department of Physics, Harvard University, Cambridge,
MA 02138, USA}

\date{June 21, 2010}

\begin{abstract}
We show that there is a close correspondence between the physical properties of
holographic metals near charged black holes in anti-de Sitter (AdS) space, and the
fractionalized Fermi liquid phase of the lattice Anderson model. The latter phase has a 
`small' Fermi surface of conduction electrons, along with a spin liquid of local moments.
This correspondence implies that certain mean-field gapless spin liquids are  
states of matter at non-zero density
realizing the near-horizon, AdS$_2 \times$R$^2$ physics
of Reissner-Nordstr\"om black holes.
\end{abstract}
\pacs{} \maketitle

There has been a flurry of recent activity \cite{Lee,Cubrovic,Liu,Faulkner,denef,tong,FP,mfl,larsen,elias} 
on the holographic description of metallic states
of non-zero density quantum matter. The strategy is to begin with a strongly interacting
conformal field theory (CFT) in the ultraviolet (UV), which has a dual description as the boundary of
a theory of gravity in anti-de Sitter (AdS) space. This CFT is then perturbed by a chemical potential ($\mu$)
conjugate to a globally conserved charge, and the infrared (IR) physics is given a holographic description
by the gravity theory. For large temperatures $T \gg \mu$, such an approach is under good control,
and has produced a useful hydrodynamic description of the physics of quantum criticality \cite{Hartnoll}. Much less
is understood about the low temperature limit $T \ll \mu$: a direct solution of the classical gravity theory
yields boundary correlation functions describing a non-Fermi liquid metal \cite{Faulkner}, but the physical interpretation
of this state has remained obscure. It has a non-zero entropy density as $T \rightarrow 0$, and this raises
concerns about its ultimate stability. 

This paper will show that there is a close parallel between the above theories of holographic
metals, and a class of mean-field theories of the `fractionalized Fermi liquid' (FFL) phase
of the lattice Anderson model. 

The Anderson model (specified below) has been a popular
description of inter-metallic transition metal or rare-earth compounds: it describes itinerant conduction
electrons interacting with localized resonant states representing $d$ (or $f$) orbitals.
The FFL is an exotic phase of the Anderson model,
demonstrated to be generically stable in Refs.~\cite{ffl1,ffl2}: it has a `small' Fermi surface whose volume is determined
by the density of conduction electrons alone, while the $d$ electrons form a fractionalized spin liquid state.
The FFL was also found in a large spatial dimension mean field theory by Burdin {\em et al.}~\cite{bgg},
and is the ground state needed for a true ``orbital-selective Mott transition'' \cite{vojta}. 
The FFL should be contrasted from the conventional Fermi liquid phase (FL), in which there is a `large' Fermi surface
whose volume counts both the conduction and $d$ electrons: the FL phase is the accepted
description of many `heavy fermion' rare-earth intermetallics.
However, recent experiments on YbRh$_2$(Si$_{0.95}$Ge$_{0.05}$)$_2$ have observed an
unusual phase for which the FFL is an attractive candidate \cite{paschen}. 

Here, we will describe the spin liquid
of the FFL by the gapless mean-field state of Sachdev and Ye \cite{sy} (SY). We will then find that
physical properties of the FFL are essentially identical to those of the present theories of holographic
metals. Similar comments apply to other gapless quantum liquids \cite{si} which are related
to the SY state.
This agreement implies that non-zero density matter described by the SY (or a related) state is a realization of 
the near-horizon, AdS$_2 \times$R$^2$ physics
of Reissner-Nordstr\"om black holes. 

We begin with a review of key features of the present theory of holographic metals. The UV physics is 
holographically described by a Reissner-Nordstr\"om black hole in AdS$_4$. In the IR, the low energy physics
is captured by the near-horizon region of the black hole, which has a AdS$_2 \times$R$^2$ geometry \cite{Faulkner}.
The UV theory can be written as a SU($N_c$) gauge theory, but we will only use gauge-invariant
operators to describe the IR physics.
We use a suggestive condensed matter notation to represent the IR, 
anticipating the correspondence we
make later.
We probe this physics by a `conduction electron' $c_{{\bf k} \alpha}$ (where ${\bf k}$ is a momentum and $\alpha = \uparrow, \downarrow$ a spin index), which will turn out to have a Fermi surface at a momentum $k \equiv |{\bf k}| = k_F$.
The IR physics of this conduction electron is described by the effective Hamiltonian \cite{Faulkner,FP}
\beqn H &=& H_0 + H_1 [\mathfrak{d},c] + H_{\rm AdS}  \label{H} \\
H_0 &=& \sum_{\alpha} \int \frac{d^2 k}{4 \pi^2} (\varepsilon_{\bf k} - \mu) c_{{\bf k} \alpha}^\dagger c_{{\bf k} \alpha},
\label{h0}
\eeqn
with $c_{{\bf k} \alpha}$ canonical fermions and $\varepsilon_{{\bf k}}$ their dispersion, and
\beq H_1 [\mathfrak{d}, c] = \sum_{\alpha} \int \frac{d^2 k}{4 \pi^2} \left[ V_{{\bf k}} \mathfrak{d}^{\dagger}_{{\bf k}\alpha} c_{{\bf k} \alpha} 
+ V_{{\bf k}}^\ast c^{\dagger}_{{\bf k}\alpha} \mathfrak{d}_{{\bf k} \alpha} \right], \label{h1}
\eeq
with $V_{{\bf k}}$ a `hybridization' matrix element. The $\mathfrak{d}_{{\bf k}\alpha}$ are non-trivial operators controlled by 
the strongly-coupled IR CFT associated with the AdS$_2$ geometry, and described by $H_{\rm AdS}$;
their long imaginary-time ($\tau$) correlation under $H_{\rm AdS}$ is given by \cite{Faulkner,FP,Iqbal} (for
$0 < \tau < 1/T$)
\beq \left\langle \mathfrak{d}_{{\bf k}\alpha} (\tau) \mathfrak{d}^\dagger_{{\bf k}\beta} (0) \right\rangle_{H_{\rm AdS}} \sim 
\left[\frac{\pi T}{\sin (\pi T \tau)} \right]^{2 \Delta_k} , \label{dd} \eeq
where $\Delta_k$ is the scaling dimension of $\mathfrak{d}_{{\bf k}\alpha}$ in the IR CFT. The $T>0$ functional form
in \Eq{dd} is dictated by conformal invariance. This $\mathfrak{d}_{{\bf k}\alpha}$ correlator implies a singular
self-energy for the conduction electrons; after accounting for it, many aspects of `strange metal' 
phenomenology can be obtained \cite{mfl}. The marginal Fermi liquid phenomenology \cite{varma} is obtained
for $\Delta_k = 1$.

The important characteristics of the above holographic description of metals, which we will need below, are: 
({\em i\/}) a conduction electron
self-energy which has no singular dependence on $k-k_F$, ({\em ii\/}) a dependence of the self energy on frequency
($\omega$) and $T$ which has a conformal form (obtained by a Fourier transform of \Eq{dd}), and ({\em iii\/})
a non-zero ground state entropy associated with the AdS$_2 \times$R$^2$ geometry.

Let us now turn to the lattice Anderson model. To emphasize the correspondence to the holographic theory, 
we continue to use $c_{{\bf k}\alpha}$ for the conduction electrons, while $d_{{\bf k} \alpha}$ are
canonical fermions representing the $d$ orbitals (these will be connected to the $\mathfrak{d}_{{\bf k} \alpha}$ below). 
Then the Hamiltonian is $H_A = H_0 + H_1 [d,c] + H_U$,
where the first two terms are still specified by Eqs.~(\ref{h0}) and (\ref{h1}), and
\beqn
H_U &=& \sum_i \left[ U n_{di\uparrow} n_{di\downarrow} + \left(\varepsilon_d - U/2 - \mu \right) d_{i \alpha}^\dagger d_{i \alpha} \right] \nn 
&~&~~~~~~- \sum_{i\neq j} t_{ij} d^\dagger_{i \alpha} d_{j \alpha} ,
\eeqn
where $d_{i\alpha}$ is the Fourier transform of $d_{{\bf k} \alpha}$ on the lattice sites $i$ at spatial positions
${\bf r}_i$
with $d_{i \alpha} = \int_k d_{{\bf k}\alpha} e^{i {\bf k} \cdot {\bf r}_i}$, $n_{di\alpha} = d_{i\alpha}^\dagger
d_{i \alpha}$ is the $d$ number operator, and $t_{ij}$ are hopping matrix elements for the $d$ electrons. 
We consider $H_A$ as the UV theory of the lattice Anderson model; it clearly differs
greatly from the UV AdS$_4$ SU($N_c$) CFT considered above. We will now show that, under suitable conditions,
both theories have the same IR limit. 

We need to study the IR limit of $H_A$ to establish this claim. We work in the limit of 
$U$ larger than all other parameters, when the occupation number of each $d$ site is unity.
As is well-known \cite{pwa,sw}, to leading order in the $t_{ij}$ and $V_{{\bf k}}$, we can eliminate the 
coupling to the doubly-occupied and empty $d$ sites by a canonical transformation $\mathcal{U}$,
and derive an effective low-energy description in terms of a Kondo-Heisenberg Hamiltonian.
Thus $H_A \rightarrow \mathcal{U} H_A \mathcal{U}^{-1}$, where the
$d_{i\alpha}$  are now mapped as $\mathfrak{d}_{i\alpha} = \mathcal{U} d_{i \alpha} \mathcal{U}^{-1}$
which yields \cite{sw,fisher}
\beqn
H_A &=& H_0 + H_1 [\mathfrak{d},c] + H_J \label{HA} \\
\mathfrak{d}_{i\alpha} 
&=& \frac{\sigma^a_{\alpha \beta}}{2} \int \frac{d^2 k}{4 \pi^2} \left[  \frac{U V_{{\bf k}} e^{- i {\bf k} \cdot {\bf r}_i}}{U^2/ 4 - (\varepsilon_d - \varepsilon_{\bf k})^2} \right]  c_{{\bf k} \beta}\, \hat{S}^a_i . \label{swt}
\eeqn
Here $\sigma^a$ ($a=x,y,z$) are the Pauli matrices 
and the $\hat{S}^a_i$ are operators measuring the spin of the $d$ local moment on site $i$. 
The $\hat{S}^a_i$ operators should be considered as abstract operators acting on the local moments:
they are fully defined by the commutation relations $[\hat{S}^a_i , \hat{S}^b_j] = i \epsilon^{abc} \delta_{ij} \hat{S}^c_i$
and the length constraint $\sum_a \hat{S}^{a2}_i = 3/4$. 
The IR physics directly involves
only the metallic $c_{{\bf k}\alpha}$ fermions (which remain canonical), and the spin
operators $\hat{S}^a_i$. The Schrieffer-Wolff transformation \cite{sw} implies that $\mathfrak{d}_{{\bf k} \alpha}$ is a composite
of these two low-energy (and gauge-invariant) operators, and is not a canonical fermion.
The canonical transformation $\mathcal{U}$ also generates
a direct coupling between the $\hat{S}^a_i$ which is
\beq
H_J = \sum_{i<j} J_{ij} \hat{S}^a_i \hat{S}^a_j \label{hj}
\eeq
where $J_{ij} = 4 |t_{ij}|^2/U$. Also note that after substituting \Eq{swt} into $H_1$ we obtain the Kondo
exchange between the conduction electron and the localized spins: here, we have reinterpreted
this Kondo interaction as the projection of the $d$ electron to the IR via \Eq{swt}.

More generally, we can view the correspondence $\mathfrak{d} \sim c\, \hat{S}$ in \Eq{swt} as the
simplest operator representation consistent with global conservation laws. We need an operator in the IR
theory which carries both the electron charge and spin $S=1/2$. The only simpler correspondence is
$\mathfrak{d} \sim c$, but this can be reabsorbed into a renormalization of the $c$ dispersion.

We now focus on the FFL phase of $H_A$ in \Eq{HA}. In this phase the influence of $H_1$ can be 
treated perturbatively \cite{ffl1} in $V_{\bf k}$, and so we can initially neglect $H_1$.
Then the $c_{{\bf k}\alpha}$ form a `small' Fermi surface defined by $H_0$, and the spins of $H_J$
are required \cite{ffl1} to form a spin liquid. As discussed earlier, we assume that $H_J$ realizes the
SY gapless spin liquid state. Such a state was formally justified \cite{sy} in the quantum analog of the
Sherrington-Kirkpatrick model, in which all the
$J_{ij}$ are infinite-range, independent Gaussian random variables with variance $J^2/N_s$
($N_s$ is the number of sites, $i$). However, it has also been shown \cite{bgg,rahul} that 
closely related mean-field equations apply to frustrated antiferromagnets with non-random exchange interactions
in the limit of large spatial dimension \cite{qmsi,chitra}. We will work here with the SY equations as the simplest representative
of a class that realize gapless spin liquids.
The SY state of $H_J$ is described by a
single-site action $\mathcal{S}$, describing the self-consistent quantum fluctuation of the spin $\hat{S}^a (\tau)$
in imaginary time. We express the spin in terms of a unit-length vector ${n}^a (\tau) = 2 \hat{S}^a (\tau) $
and then we obtain the coherent state path integral
\beqn
\mathcal{Z} &=& \int \mathcal{D}  n^a (\tau) \,  \delta ( n^{a2} (\tau) - 1) \exp \left ( - \mathcal{S} \right) \label{sy1} \\
\mathcal{S} &=& \frac{i}{2} \int d \tau \mathcal{A}^a \frac{d n^a}{d \tau} - \frac{J^2}{2} 
\int d \tau d \tau' Q (\tau - \tau') n^a (\tau) n^a (\tau'). \nonumber
\eeqn
The first term in $\mathcal{S}$ is the spin Berry phase, with $\mathcal{A}^a$ any function of $n^a$ obeying
$\epsilon^{abc} (\partial \mathcal{A}^b / \partial n^c) = n^a$. The function $Q$ is to determined self-consistently
by the solution of 
\beqn
Q(\tau - \tau') = \left\langle n^a (\tau) n^a (\tau') \right\rangle_{\mathcal{Z}}. \label{sy2}
\eeqn

The equations (\ref{sy1}) and (\ref{sy2}) define a strong-coupling problem for which no complete solution
is known. However, these equations have been extensively studied \cite{sy,pgks,pg,pgs} 
in the framework of a $1/N$ expansion
in which the SU(2) spins are generalized to SU($N$) spins, and some scaling dimensions are
believed to be known to all orders in $1/N$ \cite{pgs}. Note that the SU($N$) is a global
`flavor' symmetry.
For the SU($N$) case, we can consider general
spin representations described by rectangular Young tableaux with $m$ columns and $qN$ rows.
For such spins, the generators of SU($N$), $\hat{S}_{\alpha}^{\beta}$, (now $\alpha,\beta = 1 \ldots N$)
can be written in terms of `slave' fermions $f^{s \alpha}$ (with $s = 1 \ldots m$) by \cite{rs0}
$\hat{S}_{\alpha}^{\beta} = \sum_{s=1}^m f_{s \alpha}^\dagger f^{s \beta}$ along with the constraint
$\sum_{\alpha=1}^N f_{s \alpha}^\dagger f^{s' \alpha} = \delta_s^{s'} q  N$. When expressed in terms
of such fermions, the original lattice model $H_J$ defines a U($m$) gauge theory \cite{rs0}. 
It is worth emphasizing that the $f_{s \alpha}$ are the {\em only\/} gauge-dependent operators
considered in this paper, and the U($m$) gauge transformation acts on the $s$ index.
For $\mathcal{Z}$ in Eq.~(\ref{sy1}), the slave fermion representation enables a solution
in the limit of large $N$, at fixed $q$ and $m$.
Remarkably, the IR limit of this solution has 
the structure of a conformally-invariant (0+1)-dimensional boundary of a 1+1 dimensional CFT \cite{pgks,pg}.
In particular, for the fermion Green's function 
$G_f (\tau) = \left\langle f^{s \alpha} (\tau) f^\dagger_{s \alpha} (0) \right\rangle$ we find the conformal form \cite{sy,pg,pgs}
\beq
G_f ( \tau ) \sim \left[ \frac{\pi T}{\sin (\pi T \tau)} \right]^{1/2}. \label{Gf}
\eeq
In the large $N$ limit, $Q(\tau) \propto G_f (\tau) G_f (- \tau)$, and therefore
\beq
Q(\tau) \sim \frac{\pi T}{\sin (\pi T \tau)}. \label{Qt}
\eeq
This implies the non-trivial result that the scaling dimension of the spin operator $\hat{S}_\beta^{\alpha}$ is 1/2.
It has been argued that this scaling dimension holds to all orders in $1/N$ \cite{pgs,cb}, and so for
SU(2) we also expect $\mbox{dim}[\hat{S}^a] = 1/2$.
Other mean-field theories of $H_A$ have been studied \cite{si,qmsi,rahul,pgks,cb}, and yield related gapless quantum liquids
with other scaling dimensions, although in most cases the solution obeys the self-consistency condition
in \Eq{sy2} only with the exponent in \Eq{Qt}.

With the knowledge of \Eq{Qt}, we can now compute the physical properies of the FFL phase of $H_A$ 
associated with
the SY state. These can be computed perturbatively in $V_{{\bf k}}$, as was discussed
by Burdin {\em et al.} \cite{bgg}. They reproduced 
much of the `marginal Fermi liquid' phenomenology of Ref.~\cite{varma}, including the linear-$T$ resistivity. Note that no exponent was adjusted to achieve this (unlike Ref.~\cite{mfl});
the linear resistivity is a direct consequence of the scaling dimension sdim[$\hat{S}^a$] = 1/2. 

We are now in a position to compare the IR limit of the theory of holographic metals to the FFL phase
of $H_A$ associated with \Eq{Qt}: \newline
({\em i\/}) For $H_A$, we can easily compute the two-point
$d_{{\bf k}\alpha}$ correlator from \Eq{swt} as a product of the $c_{{\bf k}\alpha}$
and $\hat{S}^a$ correlators. For the latter, we use
 (\ref{Qt}) for the on-site correlation, and drop the off-site correlations
 which average to zero in the SY state (and in large dimension limits); it is this
 limit which leads to the absence of a singular ${\bf k}$-dependence in the $d_{{\bf k} \alpha}$ correlator. 
  For the electron, we use
the Fermi liquid result
\beq
\left\langle c_{i \alpha} ( \tau) c_{i \alpha}^\dagger (0) \right\rangle_{H_0} \sim \frac{\pi T}{\sin (\pi T \tau)},
\eeq
and then we find that the $d_{{\bf k} \alpha}$ 
correlator has the form of the holographic result in \Eq{dd} with $\Delta_k = 1$. As expected from the results for $H_A$,
this is the value of $\Delta_k$ corresponding to the marginal Fermi liquid \cite{mfl}.
\newline
({\em ii\/}) The SY state 
has a finite entropy density at $T=0$. This entropy has been computed in the large $N$
limit \cite{pgs}, and the results agree well with considerations based upon the boundary entropy
of 1+1 dimensional CFTs \cite{pgks}. The holographic metal has
also a finite entropy density, associated with the horizon of the extremal black hole.
However, a quantitative comparison of the entropies of these two states is not yet possible.
The entropy of the SY state is quantitatively 
computed \cite{pgs} in the limits of large $N$ (where SU($N$) is a flavor
group) and large spatial dimension, but at fixed $m$ and $q$. 
In contrast, the holographic metal computation is in the limit of large $N_c$
(where SU($N_c$) is the gauge group).

The above correspondences in the IR limit of the electron correlations and the thermodynamics support
our main claim that the SY-like spin liquids realize the physics of AdS$_2 \times$R$^2$. 

It is interesting to compare our arguments with the recent results of Kachru
{\em et al.} \cite{sho}. They used an intersecting D-brane construction to introduce point-like
impurities with spin degrees of freedom which were coupled to a background CFT. 
For each such impurity there was an asymptotic AdS$_2$ and an associated degeneracy
of the ground state; a lattice of impurities led to a non-zero entropy density at $T=0$. 
Thus working from their picture, it is very natural to associate 
AdS$_2 \times$R$^2$
with a lattice of interacting spins; with supersymmetry \cite{sho}, or in a mean-field theory \cite{sy},
such a model can have a non-zero entropy density. The similarity between these theories leads us 
to conjecture 
that a possible true ground state of the quantum gravity theory of the holographic 
metal is a spontaneously formed 
crystal of spins coupled to the Fermi surface of conduction electrons.
This would then be an example of quantum ``order-from-disorder'' \cite{henley}, with the
quantum ground state having a lower translational symmetry than that of the classical gravity theory.

Below, we accept our main claim connecting the holographic
metal to the FFL phase with a SY-like spin liquid, and discuss further implications.

From the perspective of $H_A$, it is not likely that the SY state is stable beyond its
large spatial dimension limit \cite{pgs}; the $d_{{\bf k} \alpha}$ 
propagator should acquire a singular ${\bf k}$-dependence in finite dimensions.
However, the remarkable emergence of the large dimension SY state in the very 
different holographic context suggests a certain robustness, and so perhaps it should
be taken seriously as a description over a wide range of intermediate energy scales.
Ultimately, it is believed that at sufficiently low energies we must cross over to a
gauge-theoretic description of a stable spin liquid with zero ground state entropy density \cite{ffl1,ffl2}.
Associated with this stable spin liquid would be a stable FFL phase in finite dimension,
whose ultimate IR structure was described earlier \cite{ffl1,ffl2}. 
It is clearly of interest to find the parallel instabilities of the 
holographic metal on AdS$_2 \times$R$^2$. The geometry should acquire corrections which
are compatible with a ${\bf k}$-dependent self energy, and this should ultimately lift the 
ground state entropy. Some of the considerations of Refs.~\cite{tong,elias} may already represent
progress in this discussion.

Refs.~\cite{ffl1,ffl2} also discussed the nature of the quantum phase transition between
the FFL and FL phases. It was argued that this was a Higgs transition which quenched
gauge excitations of the FFL spin liquid. Consequently, we conclude that a
holographic Fermi liquid can be obtained by a Higgs transition in the boundary theory.
In string theory, the Higgs transition involves separation of co-incident D-branes, and it
would be useful to investigate such a scenario here. The transition from FFL to FL involves
an expansion in the size of the Fermi surface from `small' to `large', so that the Fermi surface
volume accounts for all the fermionic matter. It is no longer permissible to work perturbatively in $V_{{\bf k}}$ in the
FL phase: instead we have to renormalize the band structure to obtain quasiparticles
that have both a $c_{{\bf k} \alpha}$ and a $d_{{\bf k} \alpha}$ character.
Present theories
of holographic metals have an extra `bath' of matter outside the Fermi surface which can
be accounted for perturbatively in $V_{{\bf k}}$---indeed, these were key reasons for identifying
them with the FFL phase. It would be interesting to obtain the Fermi surface expansion
in the holographic context.

I thank all the participants of TASI 2010, Boulder, Colorado for stimulating discussions,
and especially K.~Balasubramanian, T.~Grover, N.~Iqbal, S.-S.~Lee, H.~Liu, J.~Polchinski, and S.~Yaida.
I also thank A.~Georges (for many discussions on large dimensions and spin liquids in past years), S.~Hartnoll,
S.~Kachru, and J.~Zaanen.
This research was supported by the National Science Foundation under grant DMR-0757145, by the FQXi
foundation, and by a MURI grant from AFOSR.


\begin{thebibliography}{}

\bibitem{Lee} S.-S.~Lee, Phys Rev D {\bf 79}, 086006 (2009).

\bibitem{Cubrovic} M.~Cubrovic, J.~Zaanen and K.~Schalm,
  %``Fermions and the AdS/CFT correspondence: quantum phase transitions and the
  %emergent Fermi-liquid,''
  Science {\bf 325} 439 (2009).

\bibitem{Liu}
  H.~Liu, J.~McGreevy and D.~Vegh, arXiv:0903.2477.

\bibitem{Faulkner}
  T.~Faulkner, H.~Liu, J.~McGreevy and D.~Vegh, arXiv:0907.2694.

\bibitem{denef} F. Denef, S. A. Hartnoll, and S. Sachdev, Phys. Rev. D {\bf 80}, 126016 (2009).

\bibitem{tong} S.~A.~Hartnoll, J.~Polchinski, E.~Silverstein, and D.~Tong,
JHEP {\bf 1004}, 120 (2010).

\bibitem{FP} T.~Faulkner and J.~Polchinski,
  %``Semi-Holographic Fermi Liquids,''
  arXiv:1001.5049.

\bibitem{mfl} T.~Faulkner, N.~Iqbal, H.~Liu, J.~McGreevy, and D.~Vegh, 
Science Online, DOI: 10.1126/science.1189134

\bibitem{larsen} F.~Larsen, and G.~van Anders, arXiv:1006.1846.

\bibitem{elias} C.~Charmousis, B.~Gouteraux, B. ~S.~Kim, E.~Kiritsis, and R.~Meyer,
arXiv:1005.4690.

\bibitem{Hartnoll}
  S.~A.~Hartnoll, P.~K.~Kovtun, M.~M\"uller and S.~Sachdev,
  Phys.\ Rev.\  B {\bf 76}, 144502 (2007).

\bibitem{ffl1} T.~Senthil, S.~Sachdev, and M.~Vojta, Phys. Rev. Lett. {\bf 90}, 216403 (2003).

\bibitem{ffl2} T.~Senthil, M.~Vojta, and S.~Sachdev, Phys. Rev. B {\bf 69}, 035111 (2004).

\bibitem{bgg} S. Burdin, D. R. Grempel, and A. Georges, Phys. Rev. B {\bf 66}, 045111 (2002).

\bibitem{vojta} M.~Vojta, arXiv:1006.1559; V.~I.~Anisimov,  
I.~A.~Nekrasov, D.~E.~Kondakov, T.~M.~Rice, and M.~Sigrist, Eur. Phys. J. B {\bf 25}, 191
(2002); L.~De Leo, M.~Civelli, and G.~Kotliar, Phys. Rev. Lett. {\bf 101}, 256404 (2008).

\bibitem{paschen} J.~Custers {\em et al.},
%P.~Gegenwart, C.~Geibel, F.~Steglich, P.~Coleman, 
%and S. Paschen, 
Phys. Rev. Lett. {\bf 104}, 186402 (2010). 

\bibitem{sy} S.~Sachdev and J.~Ye, Phys. Rev. Lett. {\bf 70}, 3339 (1993).
 
\bibitem{si} L.~Zhu and Q.~Si, Phys. Rev. B {\bf 66}, 024426 (2002); L.~Zhu, S.~Kirchner, Q.~Si, and A.~Georges,
Phys. Rev. Lett. {\bf 93}, 267201 (2004).

\bibitem{Iqbal} N.~Iqbal and H.~Liu, Fortschr. Phys. {\bf 57}, 367 (2009).

\bibitem{varma} C.~M.~Varma, P.~B.~Littlewood, S.~Schmitt-Rink, E.~Abrahams 
and A.~E.~Ruckenstein, Phys. Rev. Lett. {\bf 63}, 1996 (1989).

\bibitem{pwa} P.~W.~Anderson, Phys. Rev. {\bf 79}, 350 (1950).

\bibitem{sw} J.~R.~Schrieffer and P.~A.~Wolff, Phys. Rev. {\bf 149}, 491 (1966).
Our \Eq{swt} is obtained by using their Eq. (6) for $\mathcal{U}$, and computing
$\mathcal{U} d_{i \alpha} \mathcal{U}^{-1}$.

\bibitem{fisher} G.~Moeller, Q.~Si, G.~Kotliar, M.~Rozenberg, and D.~S.~Fisher, Phys. Rev. Lett.
{\bf 74}, 2082 (1995).

\bibitem{rahul} A.~Georges, R.~Siddharthan, and S.~Florens, Phys. Rev. Lett. {\bf 87}, 277203 (2001).

\bibitem{qmsi} 
J.~L.~Smith and Q.~Si, Phys. Rev. B {\bf 61}, 5184 (2000).

\bibitem{chitra} R. Chitra and G. Kotliar, 
Phys. Rev. Lett. {\bf 84}, 3678 (2000);
Phys. Rev. B {\bf 63}, 115110 (2001).


\bibitem{pgks} O.~Parcollet, A.~Georges, G.~Kotliar, and A.~Sengupta, Phys. Rev. B {\bf 58}, 3794 (1998).

\bibitem{pg} O.~Parcollet and A.~Georges, Phys. Rev. B {\bf 59}, 5341 (1999).

\bibitem{pgs} A.~Georges, O.~Parcollet, and S.~Sachdev, Phys. Rev. B {\bf 63}, 134406 (2001).

\bibitem{cb} M.~Vojta, C.~Buragohain, and S.~Sachdev, Phys. Rev. B {\bf 61},
15152 (2000).


\bibitem{rs0} N.~Read and S.~Sachdev, Nucl. Phys. B {\bf 316}, 609 (1989).

\bibitem{sho} S.~Kachru, A.~Karch, and S.~Yaida, Phys. Rev. D {\bf 81}, 026007 (2010).

\bibitem{henley} E.~F.~Shender, Zh. Eksp. Teor. Fiz. {\bf 83}, 326 (1982) 
[Sov. Phys. JETP {\bf 56}, 178 (1982)]; C.~L.~Henley, Phys. Rev. Lett. {\bf 62}, 2056 (1989).

\end{thebibliography}
\end{document}